\documentclass{article}
\usepackage[utf8]{inputenc}
\usepackage{geometry}
\usepackage{auxlib}
\usepackage{authblk}

\usepackage{natbib}

\newcommand{\opt}{\mathrm{OPT}}

\newcommand{\intx}{x^{\mathrm{int}}}

\newcommand{\Itrc}{I^{\mathrm{trc}}}
\newcommand{\ctrc}{c^{\mathrm{trc}}}
\newcommand{\ltrc}{\ell^{\mathrm{trc}}}
\newcommand{\Btrc}{B^{\mathrm{trc}}}
\newcommand{\Etrc}{E^{\mathrm{trc}}}

\newcommand{\opttrc}{\opt^{\mathrm{trc}}}

\newcommand{\partition}{\texttt{PartMatroid}}

\title{Fair Submodular Maximization over a Knapsack Constraint}

\author[1]{Lijun Li}
\author[2]{Chenyang Xu}
\author[2]{Liuyi Yang}
\author[3]{Ruilong Zhang}

\affil[1]{\small Department of Computer Science, City University of Hong Kong, Hong Kong, China}
\affil[2]{\small Software Engineering Institute, East China Normal University, Shanghai, China}
\affil[3]{\small Department of Mathematics, Technical University of Munich, Munich, Germany}

\affil[ ]{\texttt{lijunli3-c@my.cityu.edu.hk,
		cyxu@sei.ecnu.edu.cn,
		10225101401@stu.ecnu.edu.cn,
		ruilong.zhang@tum.de}}

\date{}

\date{}

\begin{document}
	
	\maketitle

	\begin{abstract}
		We consider fairness in submodular maximization subject to a knapsack constraint, a fundamental problem with various applications in economics, machine learning, and data mining. In the model, we are given a set of ground elements, each associated with a weight and a color, and a monotone submodular function defined over them. The goal is to maximize the submodular function while guaranteeing that the total weight does not exceed a specified budget (the knapsack constraint) and that the number of elements selected for each color falls within a designated range (the fairness constraint).

While there exists some recent literature on this topic, the existence of a non-trivial approximation for the problem -- without relaxing either the knapsack or fairness constraints -- remains a challenging open question. This paper makes progress in this direction. We demonstrate that when the number of colors is constant, there exists a polynomial-time algorithm that achieves a constant approximation with high probability. Additionally, we show that if either the knapsack or fairness constraint is relaxed only to require expected satisfaction, a tight approximation ratio of $(1-1/e-\epsilon)$ can be obtained in expectation for any $\epsilon >0$.
	\end{abstract}
	
	\newpage
	\section{Introduction}

Submodular maximization is a fundamental problem in artificial intelligence and computer science, with continuous research since the 1970s~\cite{cornuejols1977exceptional,nemhauser1978best}. 
The problem involves selecting a subset of elements from a given set to maximize a submodular function defined over them. 
It captures a wide range of tasks across various domains, such as clustering~\cite{cluster1/2020,cluster2/2019,cluster3/2017}, feature selection~\cite{fs1/2021,fs2/2022,fs3/2016}, movie recommendation~\cite{mr1/2019,mr2/2022,mr3/2022}, and so on.
However, recent studies~\cite{or1/2018,icml/HFNTT23} have found that in some data mining applications, traditional submodular optimization algorithms may face a fairness issue. 
The elements in the practical dataset often come from different groups (e.g., people of varying ages or genders~\cite{icml/HFNTT23,nips/HMNTT20,fairness1/2021}), but traditional algorithms do not account for this factor, leading to an imbalance in the number of selected elements from each group.

To address this issue,~\cite{or1/2018} introduces a \emph{group fairness} notion into submodular maximization. 
Specifically, assuming that all given elements are partitioned into several groups, a solution is considered fair if the number of selected elements from each group falls within a specified range. 
One of their results is incorporating this group fairness notion into the classic cardinality-constrained submodular maximization problem, which aims to select at most $k$ elements that satisfy the fairness constraints and maximize the submodular objective. 
They employ a well-known \emph{relax-and-round} framework~\cite{CCPV11} in submodular maximization that first applies continuous greedy to obtain a fractional solution and then designs a randomized rounding procedure to produce an integral solution. 
They show that a tight approximation ratio of $(1-1/e)$ can be obtained in expectation by a relax-and-round approach.

In this paper, we explore a generalized version of the above model. 
Note that in many applications~\cite{kempe2003maximizing,krause2008near,lin-bilmes-2011-class}, submodular maximization is often subject to a more general knapsack constraint rather than a simple cardinality limit, where each element has an associated weight, and the total weight of the selected elements must not exceed a specified budget. Therefore, we focus on optimizing fair solutions under a knapsack constraint. This general constraint introduces new challenges to fair submodular maximization. Selecting elements requires simultaneously balancing their weights, group memberships, and contributions to the submodular objective.

We notice that a recent study~\cite{kdd/CLL24} also considers this generalized model. They focus on the streaming scenario and demonstrate that if fairness constraints are allowed to be violated by a factor of $1/2$, a two-pass streaming algorithm can achieve a constant approximation. 
Furthermore, we find that directly extending the algorithm from~\cite{or1/2018} to this general problem returns a $(1-1/e) $-approximation fair solution, but at the cost of violating the knapsack constraint. 
Nevertheless, to the best of our knowledge, whether a non-trivial approximation exists when both fairness constraints and the knapsack constraint must be strictly satisfied remains an open problem.

\subsection{Problem Definition and Challenges}

The problem of \emph{fair knapsack-constrained submodular maximization} (FKSM) is formally defined as follows. There is a ground element set $E=\{e_1,\ldots,e_n\}$ and a monotone submodular\footnote{A function $f: 2^{E} \rightarrow \R$ is submodular if for all $A, B \subseteq E$ we have $f(A) + f(B) \geq f(A \cap B) + f(A \cup B)$; or equivalently $f(A \cup e) - f(A) \geq f(B \cup e) - f(B)$ for all $A \subseteq B \subseteq E$ and $e \in E$. 
The function is monotone if $f(A) \leq f(B)$ for all $A \subseteq B$.} function $f:2^{E}\rightarrow \R$ defined over them. 
Each element $e\in E$ has a weight $w_e$ and an associated color $\delta_e$. 
Let group $G_i$ represent the set of elements of color $i$, and $\cG=\{G_1,\ldots,G_k\}$ be the collection of all groups. 
Each group $G_i\in \cG$ is associated with an interval $( l_i,u_i]$. 
The goal is to select an element subset $S\subseteq E$ to maximize the value of the function $f$, while ensuring that the total weight of the selected elements does not exceed a given budget $B$, i.e., $w(S) := \sum_{e\in S} w_e \leq B$, and that the number of selected elements from each group $G_i$ lies within the range $(l_i,u_i]$, i.e., $l_i < |S\cap G_i|\leq u_i$. We call the former the knapsack constraint and the latter the fairness constraint.

We refer to a special case of our model where the weight of each element is 1, i.e., the knapsack constraint simplifies to a cardinality constraint, as F\(\overline{\text{K} }\)SM. Similarly, we refer to another special case of our model where $l_i = -\infty$, i.e., there is no lower bound on the number of elements selected from each group, as \(\overline{\text{F} }\)KSM. These two special cases represent specific simplifications of the knapsack constraint and the fairness constraints, respectively. Both of them have been extensively studied in the literature~\cite{or1/2018,aaai/GBQ23} and can achieve a $(1-1/e-\epsilon)$-approximation for any $\epsilon>0$ using the relax-and-round framework. 
The framework involves two steps: first, applying the continuous greedy algorithm to generate a fractional solution $\{ x_e\in [0,1]\}_{e\in E}$ that satisfies the constraints and achieves an approximation ratio of $(1-1/e)$; and second, employing randomized pipage rounding to produce a feasible integral solution $\{ y_e\in \{0,1\}\}_{e\in E}$.

However, things become more challenging when both constraints—the knapsack and fairness constraints—are enforced without simplification, as prior methods are no longer applicable. While the first step of continuous greedy can still generate a fractional solution that meets the constraints, the second step of randomized pipage rounding encounters feasibility issues.
The main complication with the knapsack constraint lies in the varying weights of the elements. Randomized pipage rounding can only guarantee that the number of selected elements remains unchanged (i.e., $\sum_{e} y_e = \sum_e x_e$), but cannot strictly enforce the total weight constraint. This limitation explains why the problem becomes simpler in \bksm, where all element weights are uniform.

On the other hand, although randomized pipage rounding cannot strictly enforce the total weight constraint, it does ensure that the expected total weight satisfies the constraint: $\E[\sum_{e} w_ey_e ] \leq B $ and the $y$ variables are negatively correlated. Using standard knapsack enumeration techniques~\cite{chekuri2009}, it can be shown that, with high probability (w.h.p.), $ \sum_{e} w_ey_e  \leq (1+\epsilon) B $ for any $\epsilon >0$.
This result implies that we can scale the knapsack size down by a factor of $(1+\epsilon)$ to $B'=B/(1+\epsilon)$,  run the algorithm on the scaled instance, and obtain an integral solution that w.h.p. satisfies the original knapsack constraint: $\sum_{e} w_ey_e \leq (1+\epsilon) B'=B $. This knapsack scaling approach works well for \bfsm, as the simplified fairness constraints exhibit a down-closed property and slightly reducing the knapsack size only has a minor impact on the optimal objective. However, in the general FKSM problem, reducing the knapsack size even slightly may render the instance infeasible.
For example, consider an instance where the knapsack capacity is just sufficient to satisfy the lower bounds in fairness constraints. Any further reduction in capacity would result in the absence of feasible solutions, highlighting a critical challenge in handling both knapsack and fairness constraints simultaneously in the general FKSM setting.

\subsection{Our Techniques and Results}\label{sec:results}

This work makes progress in finding a non-trivial approximation for FKSM.
To overcome the challenges mentioned above, we propose a novel technique: \emph{knapsack truncating} and combine it with the \emph{randomized weighted pipage rounding}. The knapsack truncating technique can reduce an FKSM instance $\cI$ to an \bfsm instance $\bcI$ (with the same ground elements and objective function) that satisfies the following two desirable properties:
\begin{itemize}
    \item \emph{Optimality Inheritance:} There exists an optimal solution of \(\cI\) that remains feasible for \(\bcI\).
    \item \emph{Feasibility Extension:} For any feasible solution \(\bS\) of \(\bcI\), there always exists a feasible solution \(S\) of \(\cI\) such that \( f(S) \geq f(S')/2\).
\end{itemize}
When the number of groups is constant, the reduction can be performed in polynomial time. Therefore, leveraging this technique and the algorithm for the \bfsm problem, we have the following:

\begin{restatable}{theorem}{ratio}\label{thm:knap_trun}
    Given an FKSM instance with a constant number of groups, there exists a polynomial-time algorithm that achieves an approximation ratio of $\frac{1}{2}\left(1-\frac{1}{e}\right) - \epsilon $ with probability at least $1-\frac{1}{e}-\frac{1}{e^2}$ for any $\epsilon > 0$.
\end{restatable}

The other technique is a generalization of randomized pipage rounding, referred to as \emph{randomized weighted rounding}. 
The randomized weighted rounding technique has already been applied to additive objective function cases in many previous works~\cite{aaai/0001LSVW24,jacm/GandhiKPS06}.
During this process, the weights of the elements are taken into account to guide the rounding. This method ensures that the total weight remains unchanged before and after rounding. Additionally, for each element \(e\), it guarantees that \(\E[y_e] = x_e\), where \(x_e\) represents the fractional solution, and \(y_e\) denotes the rounded solution. 
We extend this technique to the {\em submodular} function case and prove that all properties of pipage rounding (e.g., negative correlation, objective concentration) still hold in randomized weighted pipage rounding (\cref{sec:concentra}), which might be useful for other problems.
By integrating this technique with the continuous greedy method, we have the following:

\begin{restatable}{theorem}{expected}\label{thm:knap_pipage}
    Given an FKSM instance, if the fairness constraints are relaxed to be satisfied in expectation, there exists a polynomial-time algorithm that returns a solution with an expected approximation ratio of $ \left(1-\frac{1}{e}\right) - \epsilon $ for any $\epsilon > 0$. 
\end{restatable}

The above theorem shows that when the fairness constraints are relaxed to require expected satisfaction, an approximation ratio of \(1 - 1/e - \epsilon\) can be achieved in expectation. As mentioned earlier, the method from~\cite{or1/2018} can be easily extended to FKSM. It guarantees a fair solution with an expected approximation ratio of \(1 - 1/e - \epsilon\) while ensuring the knapsack constraint is satisfied in expectation (a detailed proof is provided in~\cref{sec:relaxation}).
Therefore, we can conclude that if either the knapsack or fairness constraint is relaxed to require only expected satisfaction, a tight approximation ratio of \(1 - 1/e - \epsilon\) can be achieved.

\subsection{Other Related Works}

Group fairness in submodular maximization has recently attracted considerable attention.
Various research efforts have embraced the group fairness model, suggesting fair submodular maximization algorithms that operate under a range of constraints. 
These include the matroid constraint, applied in both streaming and offline settings~\cite{icml/HFNTT23,nips/HMNTT20,aistats/HTNV24}.
Fair submodular maximization subject to a knapsack constraint is recently proposed by~\cite{kdd/CLL24} under the streaming setting.
Without fairness, the problem just aims to maximize a monotone submodular function subject to a knapsack constraint, which admits a $(1-\frac{1}{e})$-approximation algorithm~\cite{S04}.
\cite{KST09} studies submodular maximization subject to a constant number of linear constraints and gives a $(1-\frac{1}{e}-\epsilon)$ approximation algorithm.
Further, if the fairness constraint only has an upper bound, then the problem becomes the submodular maximization subject to a knapsack and partition matroid constraint.
A batch of works studies the intersection with $p$-matroid and $q$-knapsacks~\cite{siamcomp/CVZ14,jmlr/FeldmanHK23,aaai/GBQ23,stoc/LeeMNS09,icml/MBK16}.
The current best algorithm is $\Omega(1/(p+q))$-approximation, which is given by~\cite{aaai/GBQ23}.

We also notice that a closely related work is the deterministic weighted pipage rounding proposed by~\cite{jco/AS04} for the max-coverage problem with knapsack constraint.
They show that the weighted pipage rounding algorithm returns a $(1-\frac{1}{e}-\epsilon)$-approximate solution.
However, since their rounding method is deterministic, it cannot be directly applied to ensure fairness in our problem.
In our technique, a randomization procedure is introduced, which enables us to maintain the fairness constraint in expectation, and demonstrates that the produced random variables are well-concentrated.

\subsection{Roadmap}

\cref{sec:pre} states some terminology and prior results.
In \cref{sec:kt}, we introduce the knapsack truncating technique and present a constant approximation algorithm for FKSM strictly respecting knapsack and fairness constraints when the number of groups is constant.
In \cref{sec:knap-pipage}, we consider the scenario where fairness constraints are allowed to be satisfied in expectation and show that a constant approximation can be achieved by a randomized weighted pipage rounding approach. The paper finally concludes in~\cref{sec:con}.
	\section{Preliminaries}\label{sec:pre}

In this section, we introduce some terminology and prior results that will be used throughout this paper.
A fractional vector $\vx =\{x_e\in [0,1]\}_{e\in E}$ is called a fractional solution of an FKSM instance if $\vx$ is a point of the {\em fair knapsack polytope} defined as follows.
\begin{definition}[Fair Knapsack Polytope] A feasible fractional solution is a point in the following polytope $\cP_{\mathrm{FK}}$:
\[
\left\{ \vx\in[0,1]^E: \sum_{e\in E}w_ex_e \leq B;  l_i < \sum_{e\in G_i}x_e \leq u_i, \forall i\in [k]  \right\}
\]
\end{definition}

In the previous section, we mention that continuous greedy can return a fractional solution with a constant approximation. However, the given submodular function $f$ cannot directly evaluate the objective value corresponding to a fractional solution. A common approach in submodular maximization is to utilize the concept of the multi-linear extension to assess fractional solutions.

\begin{definition}[Multilinear Extension]\label{def:extension}
The multilinear extension of a submodular function $f$ is a function $F$ for $\vx =\{x_e\in [0,1]\}_{e\in E}$ where 
\[
F(x):=\E_{R\sim \cD(\vx)}[f(R)]=\sum_{R\subseteq E}\left(f(R)\prod_{e\in R}x_e\prod_{e\notin R}(1-x_e)\right)~,    
\]
where \(\cD(\vx)\) represents a probability distribution over elements, where each element is sampled independently with probability \(x_e\), and \(R\) is a random subset sampled from this distribution.
\end{definition}


It is shown by~\cite{aamas/PKL21} that the problem admits a PTAS when the objective is an additive function (i.e., $f(S)=\sum_{e\in S} f(\{e\})$ for any $S\subseteq E$).
\begin{lemma}[\cite{aamas/PKL21}]
When $f$ is an additive function, there exists an algorithm with running time $\poly(n,k,\frac{1}{\epsilon})$ that returns a solution $S$ such that (\rom{1}) $\sum_{e\in S}f(\set{e})\geq (1-\epsilon)\cdot \opt_A$; (\rom{2}) $l_i < \abs{S\cap G_i}\leq u_i$ for all $i\in[k]$, where $\opt_A$ is the optimal solution under the additive function case.
\label{lem:additive-ratio}
\end{lemma}

Therefore, using \cref{lem:additive-ratio} as a subroutine, it is shown in~\cite{CCPV11} that the continuous greedy algorithm returns a good fractional solution in polynomial time.

\begin{lemma}
Given any instance of FKSM, there is a polynomial time algorithm that returns a point $x\in\cP_{\mathrm{FK}}$ such that $F(x) \geq (1-\frac{1}{e}-\epsilon)\cdot\opt$ for any $\epsilon>0$, where $\opt$ is the optimal objective value.
\label{lem:fractional-ratio}
\end{lemma}

	\section{Knapsack Truncating}\label{sec:kt}

In this section, we introduce the knapsack truncating technique and use it to prove the following theorem:

\ratio*

\begin{algorithm*}[htb] 
\caption{Knapsack Truncating}
\label{alg:kt}
\begin{algorithmic}[1]
\Require An FKSM instance characterized by $\cI = (f, \{w_e, \delta_e\}_{e\in E}, \{ l_i,u_i \}_{i\in [k]}, B)$ and a corresponding parameter set $\{\gamma_i, \beta_i\}_{i\in [k]}$.
\Ensure A reduced \bfsm instance. 
\Statex {\color{gray} // Split each group into two groups according to parameter $\gamma_i,\beta_i$ in the reduced instance }
\For{each group index $i\in [k]$}
\For{each element $e\in G_i$}
\If{$e\notin L_i(\gamma_i)$}
\State Set its new color $\bdelta_e \gets i+k$ and weight \( \bw_e \gets w_e - \frac{w(L_i(\gamma_i))-w(L_i(\beta_i))}{\gamma_i - \beta_i} \).
\Else
\State Keep its color $\bdelta_e \gets i$ unchanged and set its new weight $\bw_e \gets 0$. 
\EndIf
\EndFor
\EndFor
\Statex {\color{gray} // Construct the upper limit for each new group }
\For{each new group index $i\in [2k]$}
\If{$i\leq k$}
\State Set $ \bu_i \gets \beta_i $. 
\Else
\State Set $\bu_i \gets \gamma_i - \beta_i$.
\EndIf
\EndFor
\Statex {\color{gray} // Construct the new budget }
\State Set the new budget: $\bB \leftarrow B - \sum_{i\in [k]} w(L_i(\gamma_i))$
\State \Return $\bcI = (f, \{\bw_e, \bdelta_e\}_{e\in E}, \{\bu_i \}_{i\in [2k]}, \bB)$.
\end{algorithmic}
\end{algorithm*}

\begin{algorithm}[htb] 
\caption{Feasibility Extension}
\label{alg:fe}
\begin{algorithmic}[1]
\Require A feasible solution $\bS$ of the reduced \bfsm instance.
\Ensure A feasible solution $S$ of the original FKSM instance. 
\For{each group index $i\in [k]$}
\State Set $S \leftarrow \bS\cap (G_i \setminus L_i(\gamma_i) ) $.
\State Add elements from $L_i(\gamma_i) $ to $S$ in increasing order of weight until $|S \cap G_i| = \gamma_i$.
\EndFor
\State Let $L(\gamma):= \bigcup_{i\in [k]} L_i(\gamma_i)$
\State \Return $\argmax\{f(S), f(L(\gamma))\}$.
\end{algorithmic}
\end{algorithm}

Given an FKSM instance, let \( S^* \) denote an optimal solution for this instance. Since this section focuses on the setting where the number of element groups is constant, through polynomial-time enumeration, we can assume that we know the following for each group \( G_i \): 

\begin{itemize}
    \item The number of elements selected from \( G_i \) by \( S^* \), i.e., the value of \( \gamma_i := |S^* \cap G_i| \).  
    \item The number of elements among the top \(\gamma_i\) smallest-weight elements in \( G_i \) that are included in \( S^* \), i.e., the value of \( \beta_i := |S^* \cap L_i(\gamma_i)| \), where \( L_i(\gamma_i) \) represents the set of the \(\gamma_i\) smallest-weight elements in \( G_i \) (with ties broken arbitrarily).
\end{itemize}

The reduction method is stated in~\cref{alg:kt}. From the description, we observe that during the reduction, both the cost of each element and the knapsack size are reduced. For elements in \(L_i(\gamma_i)\), their costs are set to 0. For each element not in \(L_i(\gamma_i)\), the cost decreases by $\frac{w(L_i(\gamma_i)) - w(L_i(\beta_i))}{\gamma_i - \beta_i}$.
This reduction amount can be interpreted as the average weight of the \(\gamma_i - \beta_i\) largest elements in the subset \(L_i(\gamma_i)\). Since \(L_i(\gamma_i)\) is defined as the set of the \(\gamma_i\) elements in \(G_i\) with the smallest weights, we always have each element's new weight in the reduced \bfsm instance \(\bw_e \geq 0\).

\paragraph{Algorithmic Intuition.} The purpose of knapsack truncating is to transform an FSMK instance into an \bfsm instance such that it can be addressed efficiently by an existing algorithm.
A very natural idea is to directly set all fairness lower bounds to zero. Clearly, in this case, the \emph{optimality inheritance} property (see its definition in~\cref{sec:results} or the lemma below) would always hold. However, the \emph{feasibility extension} property is unlikely to be satisfied, as it is possible that some weight-heavy elements have already been selected, leaving no room to add more elements to meet the fairness lower bounds. 

But we notice that things change if each group contains at least \(\gamma_i\) elements with a weight of 0. In this scenario, the fairness lower bounds are guaranteed to be met because these zero-weight elements can be selected freely. 
This suggests that to ensure feasibility extension, we must introduce zero weights into the \bfsm instance---some elements' weights need to be reduced to 0. Intuitively, reducing the weights of elements in \(L_i(\gamma_i)\) to zero should have the least impact (as \(L_i(\gamma_i)\) is the set of elements with the smallest weights). Therefore, in our algorithm, we choose to let these elements become zero-weight.
Once their weights are set to 0, the knapsack budget should be reduced accordingly by \(\sum_i w(L_i(\gamma_i))\). This adjustment can be seen as reserving space in the knapsack for elements in \(L_i(\gamma_i)\), allowing them to be freely selected later. 

However, reducing the budget harms the optimality inheritance property, as the optimal solution may contain many elements outside \(L_i(\gamma_i)\). To ensure that the optimal solution from the original instance still remains feasible under the new budget, we also need to reduce the weights of elements not in \(L_i(\gamma_i)\). 
Since FKSM and \bfsm are not equivalent, we cannot guarantee that all feasible solutions will satisfy the new budget constraint after weight reduction. Nevertheless, based on the characterization of the optimal solution provided by \(\{\gamma_i, \beta_i\}\), we can carefully design the weight reduction strategy to ensure that the optimal solution remains feasible in the new \bfsm instance.

\begin{lemma}\label{lem:kt}

Given an FKSM instance $\cI$,~\cref{alg:kt} reduces it to an \bfsm instance $\bcI$ (with the same ground elements and objective function) that satisfies the following:
\begin{enumerate}[label=(\ref{lem:kt}\ablue{\alph*}),leftmargin=*,align=left]
\item \text{Optimality Inheritance:} The optimal solution $S^*$ of \(\cI\) remains feasible for \(\bcI\).
\label{prop:lem-kt:optimal}
\item \text{Feasibility Extension:} For any feasible solution \(\bS\) of \(\bcI\), there always exists a feasible solution \(S\) of \(\cI\) such that \( f(S) \geq f(\bar{S})/2\).
\label{prop:lem-kt:feasible}
\end{enumerate}
\end{lemma}

\begin{proof}
    We begin by proving the first property. We show that \( S^* \) satisfies both the fairness and budget constraints of the reduced \bfsm instance \(\bcI\).

    For the fairness constraints, according to the description in~\cref{alg:kt}, each group \( G_i \) in the original instance $\cI$ is split into two groups: \( L_i(\gamma_i) \) and \( G_i \setminus L_i(\gamma_i) \), with corresponding fairness upper bounds of \( \beta_i \) and \( \gamma_i - \beta_i \), respectively. Due to the definitions of \( \gamma_i \) and \( \beta_i \), we directly have \( |S^* \cap L_i(\gamma_i)| = \beta_i \) and \( |S^* \cap (G_i \setminus L_i(\gamma_i))| = \gamma_i - \beta_i \). Therefore, \( S^* \) satisfies the fairness constraints of \(\bcI\).

    For the budget constraint, we have
    \begin{align*}
        \bw(S^*) =  \sum_{i\in [k]} & \bw(S^*\cap L_i(\gamma_i)) + \bw(S^*\cap (G_i \setminus L_i(\gamma_i))) \\
        =  \sum_{i\in [k]} &w(S^*\cap L_i(\gamma_i)) - w(S^*\cap L_i(\gamma_i))  \\
         & + w(S^*\cap (G_i \setminus L_i(\gamma_i)))\\
        & - (\gamma_i - \beta_i) \cdot \frac{w(L_i(\gamma_i)) - w(L_i(\beta_i))}{\gamma_i - \beta_i} \tag{Definitions of $\bw_e$, $\gamma_i$ and $\beta_i$} \\
        \leq B - &  \sum_{i\in [k]}  w(L_i(\gamma_i)) \\ 
        & + \sum_{i\in [k]} w(L_i(\beta_i)) - w(S^*\cap L_i(\gamma_i) ) \tag{Feasibility of $S^*$ for $\cI$} \\
        \leq \bB \; &,
    \end{align*}
 where the last inequality follows from the definition of $\bB$, the fact that $L_i(\beta_i)$ is the top \(\beta_i\) smallest-weight elements in $G_i$, and that $|S^*\cap L_i(\gamma_i)|=\beta_i$. Therefore, $S^*$ remains feasible for $\bcI$.

For the second property, given a feasible solution \(\bS\) for \(\bcI\), we construct a solution following the procedure described in~\cref{alg:fe}. We first prove the feasibility and then analyze its objective value. 
The algorithm ultimately selects the solution with the larger objective value between $S$ and $L(\gamma)$. Since $L(\gamma)$ is clearly feasible, the following focuses on proving the feasibility of $S$.
By the algorithm's construction, \(S\) satisfies the fairness constraints in \(\cI\), as elements are added to \(S\) until \( |S \cap G_i| = \gamma_i \) for each \(i \in [k]\). 
The proof of the budget constraint relies on the fact that elements are added in increasing order of weight (in Line 3 of~\cref{alg:fe}). Intuitively, this ensures that the average weight among any subset of these elements is at most \(\frac{w(L_i(\gamma_i)) - w(L_i(\beta_i))}{\gamma_i - \beta_i}\). For notational simplicity, let $\kappa_i := |\bS \cap (G_i \setminus L_i(\gamma_i) )|$ for each $i\in [k]$, and due to Line 3 in~\cref{alg:fe}, we have $ |S \cap  L_i(\gamma_i) | = \gamma_i -\kappa_i $.
Thus,
 \begin{align*}
        w(S) =  \sum_{i\in [k]} & w(\bS\cap (G_i \setminus L_i(\gamma_i) )) + w(S \cap  L_i(\gamma_i)) \\
        = \sum_{i\in [k]} & \bw(\bS\cap (G_i \setminus L_i(\gamma_i) )) +  w(S \cap  L_i(\gamma_i)) \\
        & +   \kappa_i \cdot \frac{w(L_i(\gamma_i))- w(L_i(\beta_i))}{\gamma_i - \beta_i}\\ 
        \leq B  & +  \sum_{i\in [k]} -w(L_i(\gamma_i)) +  w(S \cap  L_i(\gamma_i)) \\
        & +   \kappa_i \cdot \frac{w(L_i(\gamma_i))- w(L_i(\beta_i))}{\gamma_i - \beta_i}\\ \tag{Feasibility of $\bS$ for $\bcI$} \\
        \leq B  & +  \sum_{i\in [k]} -w(L_i(\gamma_i)) + w(L_i(\beta_i)) \\
         & \quad +  (\gamma_i - \kappa_i - \beta_i) \cdot \frac{w(L_i(\gamma_i))- w(L_i(\beta_i))}{\gamma_i - \beta_i} \\
        & \quad +   \kappa_i \cdot \frac{w(L_i(\gamma_i))- w(L_i(\beta_i))}{\gamma_i - \beta_i} \\
        \; = B &,
    \end{align*}
    where the last inequality uses the fact that, by the algorithm's construction, \( S \cap L(\gamma_i) \) consists of the \(\gamma_i - \kappa_i\) smallest-weight elements in \( L(\gamma_i) \).

    To prove the objective value guarantee, we observe that \( \bS \) can be partitioned into two parts: \( \bS \cap L(\gamma) \) and \( \bS \setminus L(\gamma) \), which are subsets of \( L(\gamma) \) and \( S \), respectively. Since $f$ is monotone and submodular, we can complete the proof:
    \begin{align*}
        f(\bS) & \leq f (\bS \cap L(\gamma)) + f(\bS \setminus L(\gamma) ) \tag{Submodularity} \\
        & \leq f (L(\gamma)) + f(S) \tag{Monotonicity} \\
        & \leq 2 \cdot \max\{ f (L(\gamma)), f(S) \}~.
    \end{align*}
\end{proof}

~\cref{lem:kt} demonstrates a desirable reduction from FKSM to \bfsm. The last piece in proving \cref{thm:knap_trun} is to use an existing algorithm to handle \bfsm.

\begin{lemma}[\cite{chekuri2009}]\label{lem:bfmk}
    For the \bfsm problem, there exists a polynomial time algorithm that returns an approximation ratio of $1-\frac{1}{e}-\epsilon$ with probability at least $1-\frac{1}{e}-\frac{1}{e^2}$ for any $\epsilon > 0$.
\end{lemma}

\begin{proof}[Proof of~\cref{thm:knap_trun}]

    The theorem can be easily proven using the above lemmas. We first reduce the given FKSM instance using \cref{alg:kt}, then apply the algorithm provided in \cite{chekuri2009} to the reduced \bfsm instance. Finally, we use \cref{alg:fe} to produce a feasible solution for the original problem. \cref{lem:bfmk} ensures that we can obtain a solution with an approximation ratio of \(1 - \frac{1}{e} - \epsilon\) with high probability in the reduced instance, while the two properties from \cref{lem:kt} guarantee that the solution constructed by \cref{alg:fe} will have at most a \(1/2\) decrease in its approximation ratio, which completes the proof.
\end{proof}
	\section{Randomized Weighted Pipage Rounding}
\label{sec:knap-pipage}

In this section, we apply randomized weighted pipage rounding to the submodular function case and aim to establish the following theorem.

\expected*

\begin{algorithm}[htb] 
\caption{Randomized Weighted Pipage Rounding}
\label{alg:knapsack-pipage}
\begin{algorithmic}[1]
\Require An instance with $(f, \{w_e, \delta_e\}_{e\in E}, \{ l_i,u_i \}_{i\in [k]}, B)$  and a fractional solution $\vx $ 
\Ensure A rounded solution $\vy$.
\State Initialize solution $\vy \gets \vx$.
\While{there are two elements $p,q$ with $y_p,y_q\in(0,1)$}
\State Let $\delta_1 \gets \min\{ y_p, (1-y_q)\cdot\frac{w_q}{w_p} \}$.
\State Let $\delta_2 \gets \min\{1-y_p,y_q\cdot\frac{w_q}{w_p} \}$.
\State With probability $\frac{\delta_2}{\delta_1+\delta_2}$:  
\Statex \hspace*{5em} Set $y_p \gets y_p - \delta_1$ and $y_q \gets y_q + \delta_1\cdot\frac{w_p}{w_q}$.
\State Otherwise: 
\Statex \hspace*{5em} Set $y_p \gets y_p + \delta_2$ and $y_q \gets y_q-\delta_2\cdot\frac{w_p}{w_q}$.
\EndWhile
\State \Return $\vy$.
\end{algorithmic}
\end{algorithm}

The description of the generalized rounding algorithm is provided in~\cref{alg:knapsack-pipage}.
Note that in the classic pipage rounding~\cite{CCPV11} for cardinality constraint, $\delta_1$ and $\delta_2$ are set to $\min\{ y_p, (1-y_q) \}$ and $\min\{ 1- y_p, y_q \}$, respectively. 
And, the coordinates $y_p$ and $y_q$ always change at the same rate.
However, in the weighted version, we utilize the weights of the elements to guide the rounding process, ensuring that the knapsack constraint is always satisfied. 
We begin by demonstrating several properties of weighted pipage rounding.

\subsection{Rounding Properties}

In this subsection, we discuss the properties of weighted pipage rounding from three perspectives: the knapsack constraint (\cref{lem:knap-pipage:knap}), the fairness constraints (\cref{lem:knap-pipage:fair}), and the objective guarantee (\cref{lem:knap-pipage:obj}). 

\begin{lemma}[Knapsack Satisfaction]
$\sum_{e\in E} w_ey_e \leq B$.    
\label{lem:knap-pipage:knap}
\end{lemma}

\begin{lemma}[Expected Fairness Satisfaction]
For each group $G_i$, $l_i < \E[\sum_{e\in G_i} y_e] \leq  u_i$.    
\label{lem:knap-pipage:fair}
\end{lemma}

\begin{lemma}[Objective Guarantee]
In any iteration, we have $\E[ F(\vy) ]  \geq F(\vx)$.  
\label{lem:knap-pipage:obj}
\end{lemma}

The first two lemmas can be easily proven by performing some mathematical calculations based on the probability distribution defined in~\cref{alg:knapsack-pipage}.
Due to space limitations, we defer their detailed proofs to~\cref{app:kp}. 

In the following, we focus on proving the third lemma. During the proof, we build on an existing result (\cref{lem:extension-diff}) to generalize the convexity of a submodular function's multilinear extension (\cref{lem:plane-convex}) and then use this convexity property to complete the proof.

\begin{lemma}[\cite{CCPV11}]
Let $F:[0,1]^{E}\to\R_{\geq 0}$ be the multilinear extension of a set function $f:2^{E}\to\R_{\geq 0}$. 
\begin{enumerate}[label=(\ref{lem:extension-diff}\ablue{\alph*}),leftmargin=*,align=left]
\item $\frac{\partial^2 F}{\partial x_i^2}=0$ holds at any point in $[0,1]^{E}$ for any $i\in E$. 
\label{pro:partial-square}
\item If $f$ is submodular, then $\frac{\partial^2 F}{\partial x_i\partial x_j}\leq 0$ holds at any point in $[0,1]^{E}$ for all $i,j\in E$
\label{pro:partial-diff}
\end{enumerate}
\label{lem:extension-diff}
\end{lemma}

\begin{lemma}[Convexity of Multilinear Extension]
Let \( \ve_i \) denote the vector with a value of 1 in the \( i \)-th dimension and 0 in all other dimensions. If $f$ is a monotone submodular function, then its multilinear extension is convex along the line $(c_i \ve_i- c_j \ve_j)$ for any $i,j\in E$ and constants $c_i,c_j\geq 0$, i.e., the function $F^{\va}_{ij}(\lambda):=F(\va+c_i\lambda \ve_i-c_j\lambda \ve_j)$ is a convex function for any $\va\in[0,1]^{E}, i,j\in E$ and constants $c_i,c_j
\geq 0$.
\label{lem:plane-convex}
\end{lemma}
\begin{proof}
To show the function $F^{\va}_{ij}(\cdot)$ is a convex function, it is sufficient to show the second derivative of $F^{\va}_{ij}$ is non-negative.
We define the function $\vg(\lambda):=\va+c_i\lambda \ve_i-c_j\lambda \ve_j$.
Note that $\vg(\lambda)$ is a set of functions.
For each $e\in E$, define $\vg_e(\lambda)$ as the $e$-th function of $\vg(\lambda)$, i.e., $g_e(\lambda)=a_e$ for all $e\in E\setminus\set{i,j}$, $g_i(\lambda)= a_i + c_i\lambda$ and $g_j(\lambda)=a_j-c_j\lambda$.
So, $F^{\va}_{ij}$ is a composition function: $F^{\va}_{ij}=F(g_1(\lambda),\ldots,g_n(\lambda))$.
For simplicity, let $\Delta := \va+c_i\lambda \ve_i-c_j\lambda \ve_j$. By the chain rule, we have
\[
\left.
\frac{\mathrm{d} F_{ij}^{\va}}{\partial\lambda}=\sum_{e=1}^{n}\frac{\partial F}{\partial g_i}\cdot\frac{\partial g_i}{\partial \lambda}=\left( c_i\frac{\partial F}{\partial x_i}-c_j\frac{\partial F}{\partial x_j} \right)\right\vert_{\Delta}~.
\]
Differentiating one more time and by the chain rule again, we have
\begin{align*}
&\frac{\mathrm{d}^2 F_{ij}^{\va}}{\partial\lambda^2}
=\left.\left(
c_i^2\frac{\partial^2 F}{\partial x_i^2}- 2c_ic_j \frac{\partial^2 F}{\partial x_i \partial x_j}+c_j^2\frac{\partial^2 F}{\partial x_j^2}
\right)\right\vert_{\Delta}~.
\end{align*}
Since $c_i,c_j\geq 0$ and by \cref{lem:extension-diff}, we get $\frac{\mathrm{d}^2 F_{ij}^{\va}}{\partial\lambda^2}\geq 0$.
Thus, $F(\va+c_i\lambda \ve_i-c_j\lambda \ve_j)$ is a convex function of $\lambda$.
\end{proof}

\begin{proof}[Proof of~\cref{lem:knap-pipage:obj}]
Let $\vy^{(t)}$ denote the solution $\vy$ in iteration $t$. We prove the lemma inductively by showing that $\E[F(\vy^{(t+1)})\mid \vy^{(t)}]\geq F(\vy^{(t)})$. 
Assume that $\vy^{(t+1)}$ and $\vy^{(t)}$ differ at $p,q$-th coordinates. 
Let \( \ve_q \) denote the vector with a value of 1 in the \( q \)-th dimension and 0 in all other dimensions.  
By the description of~\cref{alg:knapsack-pipage}, $\vy^{(t+1)}$ becomes \(\vy^{(t)}-\delta_1\cdot \ve_p+\delta_1\cdot \frac{w_q}{w_p}\cdot \ve_q\) with probability $\frac{\delta_2}{\delta_1+\delta_2}$, and \(\vy^{(t)}+\delta_2\cdot \ve_p-\delta_2\cdot \frac{w_q}{w_p}\cdot \ve_q\) with probability $\frac{\delta_1}{\delta_1+\delta_2}$.
Define function \(H(\lambda):=F(\vy^{(t)}+ \lambda \cdot \ve_p-\frac{w_p}{w_q} \cdot \lambda \cdot \ve_q).\)
Then, we have
\begin{align*}
\E[F(&\vy^{(t+1)})\mid \vy^{(t)}] \\
&= \frac{\delta_2}{\delta_1+\delta_2} \cdot H(-\delta_1) + \frac{\delta_1}{\delta_1+\delta_2}\cdot H(\delta_2)  \\
&\geq H \left(-\frac{\delta_2\cdot \delta_1}{\delta_1+\delta_2}+\frac{\delta_1\cdot\delta_2}{\delta_1+\delta_2}\right) \tag{\cref{lem:plane-convex}}\\
&=H(0) \\
&=F(\vy^{(t)})~.
\end{align*}
\end{proof}

\subsection{Constant Approximation with Relaxed Fairness}


This subsection proves~\cref{thm:knap_pipage}. In~\cref{lem:fractional-ratio}, we demonstrate that the continuous greedy technique can always yield a fractional solution $\vx$ with an approximation ratio of $(1 - 1/e - \epsilon)$. Subsequently, we apply~\cref{alg:knapsack-pipage} to round $\vx$ into a solution $\vy$. Using~\cref{lem:knap-pipage:knap},~\cref{lem:knap-pipage:fair} and~\cref{lem:knap-pipage:obj}, we can guarantee that $\vy$ satisfies the knapsack constraint, (expected) fairness constraints, and maintains the objective value.

However, it is worth noting that the proof does not conclude here. The solution $\vy$ is not guaranteed to be an integral solution. Due to the different rates at which $y_e$ increases or decreases during the rounding process, $\sum_{e} y_e$ may no longer be an integer. In fact, $\vy$ is a \emph{nearly} integral solution, where ``nearly'' means that at most one element has a fractional $y_e\in (0,1)$, while the remaining components are integers. 

\begin{lemma}\label{lem:nearly_interger}
    The solution $\vy$ returned by~\cref{alg:knapsack-pipage} is a nearly integral solution. Let $\vz := \{ z_e =\lfloor y_e \rfloor \}_{e\in E}$. We have integral solution $\vz$ satisfies the knapsack constraint and (expected) fairness constraint.
\end{lemma}
 \begin{proof}
     It is easy to see that $\vy$ is a nearly integral solution because if there were at least two fractional components, ~\cref{alg:knapsack-pipage} would continue iterating in the while loop. 
     Since $z_e$ is the floor of each $y_e$, it follows that $z_e \leq y_e$. By~\cref{lem:knap-pipage:knap}, we can verify the satisfaction of the knapsack constraint: 
\[\sum_{e \in E} w_e z_e \leq \sum_{e \in E} w_e y_e \leq B.\]

For the fairness constraints, the continuous greedy process ensures that the $\ell_1$ norm of the fractional solution within any group is an integer, i.e., $\sum_{e \in G_i} x_e \geq l_i + 1$. By~\cref{lem:knap-pipage:fair}, the expected $\ell_1$ norm of $\vy$ also satisfies $\E[\sum_{e \in G_i} y_e] \in  [l_i + 1,u_i]$. Furthermore, since $\vy$ is nearly integral, the rounding process decreases the $\ell_1$ norm by less than 1. Hence, for each group $G_i$, we have:
\[
\sum_{e\in G_i} y_e -1 < \sum_{e\in G_i} z_e \leq \sum_{e\in G_i} y_e \implies \E\left[\sum_{e \in G_i} z_e\right] \in (l_i, u_i].\]
 \end{proof}

\begin{proof}[Proof of~\cref{thm:knap_pipage}]

Using the above lemmas, we obtain an integer solution $\vz$ that satisfies the knapsack constraint and the expected fairness constraints. The last step is to analyze the objective value of $\vz$. According to the knapsack enumeration technique applied in~\cite{chekuri2009}, we can always ensure, through a polynomial-time enumeration process, that all elements selected via the relax-and-round method contribute only a small constant $\eta$ to the objective function value. Namely, removing any single element reduces the objective value by at most a factor of $(1+\eta)$. Therefore, $F(\vz)$ differs from $F(\vy)$ by at most a factor of $(1+\eta)$. Furthermore, ~\cref{lem:knap-pipage:obj} implies that the expected approximation ratio of $\vz$ is $1 - 1/e - \epsilon$.
\end{proof}

\subsection{Further Discussion}
\label{sec:concentra}

In this subsection, we discuss some additional properties of the randomized weighted pipage rounding, which we believe to be of independent interest. Notably, we observe that the \emph{negative correlation} and \emph{objective concentration} properties, previously established in the classic pipage rounding, still hold in the weighted version.

\begin{lemma}[Negative Correlation]
The random variables in $\set{y_e}_{e\in E}$ are negatively correlated, which means that for any $S\subseteq E$:
$
\E\left[\prod_{e\in S} y_e \right] \leq \prod_{e\in S}\E[y_e].
$    
\label{lem:negative}
\end{lemma}

\begin{lemma}[Objective Concentration]
If the objective function $f$ has a marginal value at most $1$, for any $\delta\in(0,1)$, we have $\Pr[F(\vy)\leq (1-\delta)F(\vx)]\leq\exp(-\delta^2\cdot F(\vx)/2)$.
\label{lem:concentration}
\end{lemma}


We remark that when generalized to the weighted pipage rounding, many prior properties still hold, but the original proofs are no longer applicable. Since the algorithm now moves the fractional solution along the line $\ve_p - \ve_q \cdot \frac{w_p}{w_q}$ rather than the specific direction $\ve_p - \ve_q$, all relevant proofs need to be extended. For example, when proving the objective concentration lemma, it is necessary to establish a more general property of the concave pessimistic estimator (see \ref{pro:concave} of \cref{lem:est-knap}).

	\section{Conclusion}\label{sec:con}

In this paper, we consider the problem of fair submodular maximization under a knapsack constraint. We introduce two novel techniques: knapsack truncating and randomized weighted pipage rounding, and apply them to derive fair solutions with good approximations. Several directions for future research remain open. For example, when the number of groups is non-constant, it still remains an open question if a non-trivial approximation can be achieved. Investigating whether our proposed techniques can be further extended to this more general setting, or if entirely new techniques are required to resolve this open problem, is an interesting direction for future work.

	\newpage
	\clearpage
	\bibliographystyle{plain}
	\bibliography{main}

	\newpage
	\clearpage
	\appendix
	
	\section{Expected Knapsack Satisfaction Relaxation}
\label{sec:relaxation}


In this section, we show that the classical pipage rounding satisfies the fairness constraint in expectation.
To ensure fairness inside each group, we need to apply a group pipage rounding (\cref{alg:group-pipage}).
The algorithm just modifies the updating sequence of the coordinate.
Thus, the group version has all pipage rounding properties.
We aim to show the following theorem.

\begin{theorem}
There is an algorithm with running time $\poly(n,\frac{1}{\epsilon})$ that returns an element set $S$ such that (\rom{1}) $\E[c(S)]\leq B$; (\rom{2}) $l_i\leq \abs{S\cap G_i} \leq u_i$; (\rom{3}) $\E[f(S)]\geq (1-\frac{1}{e}-\epsilon)\opt$.
\label{thm:knapsack-relax}
\end{theorem}

\begin{algorithm}[htb] 
\caption{Group Pipage Rounding \protect\cite{or1/2018} }
\label{alg:group-pipage}
\begin{algorithmic}[1]
\Require A fractional solution $\vx\in\cP_{\mathrm{FK}}$; Element set $E:=\set{e_1,\ldots,e_n}$; Groups $\cG:=\set{G_1,\ldots,G_k}$.
\Ensure An integral solution. 
\For {each group $G_i\in\cG$}
\While {there are two elements in $p,q\in G_i$ such that $x_p,x_q\in(0,1)$}
\State $\delta_1 \gets \min\{x_p,1-x_q\}$ and $\delta_2 \gets \min\{1-x_p,x_q\}$.
\State with probability $\frac{\delta_2}{\delta_1+\delta_2}$:  $x_p \gets x_p - \delta_1$ and $x_q \gets x_q + \delta_1$.
\State\hspace{5.3em} otherwise: $x_p \gets x_p + \delta_2$ and $x_q \gets x_p-\delta_2$.
\EndWhile
\EndFor
\While{there exists an element $p$ in $E$ such that $x_p\in(0,1)$}
\State Pick another fractional element $q$ in $E$.
\State $\delta_1 \gets \min\{x_p,1-x_q\}$ and $\delta_2 \gets \min\{1-x_p,x_q\}$.
\State with probability $\frac{\delta_2}{\delta_1+\delta_2}$:  $x_p \gets x_p - \delta_1$ and $x_q \gets x_q + \delta_1$.
\State\hspace{5.3em} otherwise: $x_p \gets x_p + \delta_2$ and $x_q \gets x_p-\delta_2$.
\EndWhile
\State \Return $\vx$.
\end{algorithmic}
\end{algorithm}

Let $\intx$ be the integral solution returned by \cref{alg:group-pipage}.
We define $S:=\set{e\in E: \intx_e=1}$.
Let $X_e\in\set{0,1}$ be a random variable that stands for the value of $\intx_e$.
It is well known that $\E[f(S)]\geq (1-\frac{1}{e}-\epsilon)$ and we shall not present the proof for this part.
For (\rom{1}) and (\rom{2}) of \cref{thm:knapsack-relax}, we prove them in \cref{lem:knapsack-relax:knapsack} and \cref{lem:knapsack-relax:fairness}, respectively.

\begin{lemma}
We have $\E[w(S)]\leq B$.
\label{lem:knapsack-relax:knapsack}
\end{lemma}

\begin{proof}
It is equivalent to show $\E[\sum_{e\in E}w_e X_e]\leq B$.
This is implied by $\sum_{e\in E}\E[w_e X_e] \leq B$ by the linearity of the expectation.
This is true because the marginal probability is respected by the pipage rounding algorithm, i.e., $\E[X_e]=x_e$.
Since $x$ is a point in $\cP_{\mathrm{FK}}$, we have $\sum_{e\in E} w_e x_e \leq B$.
Thus, the lemma is true.
\end{proof}

\begin{lemma}
For each $i\in[k]$, we have $l_i< \abs{S\cap G_i}\leq u_i$.
\label{lem:knapsack-relax:fairness}
\end{lemma}

\begin{proof}
We fix any group $i\in[k]$.
Observe that during the first while-loop of \cref{alg:group-pipage}, the value of $\sum_{e\in G_i}x_e$ maintains over all iterations.
At the end of the first while-loop, there is at most one fractional coordinate.
Such a coordinate may decrease to $0$ after the second while-loop of \cref{alg:group-pipage}.
Thus, either $\sum_{e\in G_i}\intx_e=\floor{\sum_{e\in G_i}x_e}$ or $\sum_{e\in G_i}\intx_e=\ceil{\sum_{e\in G_i}x_e}$ holds.
Since $x$ is a point in $\cP_{\mathrm{FK}}$, we have $l_i< \sum_{e\in G_i}x_e \leq u_i$.
Hence, we have $l_i< \sum_{e\in G_i}\intx_e \leq u_i$ since $l_i$ and $u_i$ are integers.
\end{proof}

	\section{Missing Proofs in~\cref{sec:knap-pipage}}\label{app:kp}

\subsection{Proof of~\cref{lem:knap-pipage:knap}}
Let $\vy^{(t)}$ denote the solution $\vy$ in iteration $t$.
As in the beginning, we have $\sum_{e\in E} w_ex_e \leq B$, the proof can be completed by showing that $\sum_{e\in E}w_e\cdot y^{(t)}_e = \sum_{e\in E} w_e \cdot y^{(t+1)}_e$. 
Note that $\vy^{(t+1)}$ is obtained via $\vy^{(t)}$ by updating only two coordinates, denoted by $p,q$.
Thus, it is sufficient to prove that 
\begin{equation}
w_p\cdot y^{(t)}_p + w_q \cdot y^{(t)}_q = w_p\cdot y^{(t+1)}_p + w_q \cdot y^{(t+1)}_q.   
\label{equ:keep-knapsack}
\end{equation}
We distinguish two cases: 
\begin{itemize}
    \item[(1)] $y^{(t+1)}_p=y^{(t)}_p-\delta_1$ and $y^{(t+1)}_q = y^{(t)}_q+\delta_1\cdot\frac{w_p}{w_q}$
    \item[(2)] $y^{(t+1)}_p = y^{(t)}_p+\delta_2$ and $y^{(t+1)}_q=y^{(t)}_q-\delta_2\cdot\frac{w_p}{w_q}$.
\end{itemize}

In the first case, if $y^{(t)}_p\leq (1-y^{(t)}_q)\cdot\frac{w_q}{w_p}$, then $\delta_1 = y^{(t)}_p$, implying that $y^{(t+1)}_p=y^{(t)}_p-y^{(t)}_{p}$ and $y^{(t+1)}_q = y^{(t)}_q+y^{(t)}_p\cdot\frac{w_p}{w_q}$. 
We have
\[
w_p\cdot y^{(t+1)}_p + w_q \cdot y^{(t+1)}_q = w_p\cdot y^{(t)}_p-w_p\cdot y^{(t)}_p + w_q\cdot y^{(t)}_q + w_q\cdot y^{(t)}_p\cdot \frac{w_p}{w_q}.
\]
On the other hand, if $y^{(t)}_p> (1-y^{(t)}_q)\cdot\frac{w_q}{w_p}$, then $\delta_1 = (1-y^{(t)}_q)\cdot\frac{w_q}{w_p}$, implying that $y^{(t+1)}_p=y^{(t)}_p-(1-y^{(t)}_q)\cdot \frac{w_q}{w_p}$ and $y^{(t+1)}_q = 1$.
We have
\[
w_p\cdot y^{(t+1)}_p + w_q \cdot y^{(t+1)}_q = w_p\cdot\left( y^{(t)} - \frac{w_q}{w_p}+y^{(t)}_q\cdot\frac{w_q}{w_p}\right)+w_q.
\]
Hence, \cref{equ:keep-knapsack} holds.

In the second case, similarly, if $1-y^{(t)}_p\leq y^{(t)}_q\cdot\frac{w_q}{w_p}$, then $\delta_2=1-y^{(t)}_p$, implying that $y^{(t+1)}_p=1$ and $y^{(t+1)}_q=y^{(t)}_q-(1-y^{(t)}_p)\cdot\frac{w_p}{w_q}$.
We have
\[
w_p\cdot y^{(t+1)}_p + w_q \cdot y^{(t+1)}_q = w_p+w_q\cdot\left( y^{(t)}_q-(1-y^{(t)}_p)\cdot\frac{w_p}{w_q} \right).
\]
On the other hand, if $1-y^{(t)}_p> y^{(t)}_q\cdot\frac{w_q}{w_p}$, then $\delta_2=y^{(t)}_q\cdot\frac{w_q}{w_p}$, implying that $y^{(t+1)}_p=y^{(t)}_p+y^{(t)}_q\cdot\frac{w_q}{w_p}$ and $y^{(t+1)}_q=y^{(t)}_q-y^{(t)}_q$.
We have
\[
w_p\cdot y^{(t+1)}_p + w_q \cdot y^{(t+1)}_q = w_p\cdot y^{(t)}_p + w_p\cdot y^{(t)}_q\cdot\frac{w_q}{w_p}+ w_q\cdot y^{(t)}_q-w_q\cdot w^{(t)}_q.
\]
Hence, \cref{equ:keep-knapsack} holds and the proof has been completed.

\subsection{Proof of~\cref{lem:knap-pipage:fair}}

    Notice that the fractional solution \( \vx \) satisfies the fairness constraints. We also employ an inductive argument to prove this theorem: $\E[y_e^{(t+1)}\mid \vy^{(t)}] = y_e^{(t)}$ for any iteration $t$ and element $e\in E$. 
    
    Observe that $\vy^{(t+1)}$ and $\vy^{(t)}$ differ by only two coordinates, denoted by $p$ and $q$. It is sufficient to show $\E[y^{(t+1)}_p\mid \vy^{(t)}]=y^{(t)}_p$ and $\E[y^{(t+1)}_q\mid \vy^{(t)}]=y^{(t)}_q$.
By the rounding procedure, we have
\begin{align*}
\E[y^{(t+1)}_p\mid \vy^{(t)}]
&= \frac{\delta_2}{\delta_1+\delta_2} \cdot (y^{(t)}_p-\delta_1)+\frac{\delta_1}{\delta_1+\delta_2}\cdot (y^{(t)}_p+\delta_2) \\
&=\frac{(\delta_1+\delta_2)\cdot y^{(t)}_p-\delta_1\delta_2+\delta_1\delta_2}{\delta_1+\delta_2}
=y^{(t)}_p
\end{align*}
 Similarly, the equation for the other element $q$ can also be proved.




\subsection{Proof of~\cref{lem:negative}}
Consider any set $S$ of elements and any iteration $t$.
Similarly to the proof of the negative correlation in classical pipage rounding, it is sufficient to prove the following inequality:
\[
\E\left[ \prod_{e\in S}y^{(t+1)}_e \mid \set{y^{(t)}_e}_{e\in S} \right] \leq \prod_{e\in S}y^{(t)}_e.
\]
Utilizing the inequality mentioned above, we can employ the law of total expectation and account for every iteration within $\set{0,1,\ldots,\ell-1}$. 
Consequently, we obtain that $\E\left[ \prod_{e\in S}y^{(\ell)}_e \right]\leq \E\left[ \prod_{e\in S} y^{(0)}_e \right]$.
The left-hand side is $\E\left[ \prod_{e\in S}y_e \right]$.
The right-hand side is $\prod_{e\in S}x_e$ since $y^{(0)}_e=x_e$ with probability $1$.
According to the inductive augment stated in the proof of~\cref{lem:knap-pipage:fair}, we have $\prod_{e\in S}x_e=\prod_{e\in S}\E[y^{(\ell)}_e]$, completing the proof of the negative correlation property.

Now we prove the inequality above. Let $p,q$ be the updated element during the $(t+1)$-th iteration.
We distinguish three cases: (\rom{1}) $p,q\notin S$; (\rom{2}) $\abs{S\cap\set{p,q}}=1$; (\rom{3}) $p,q\in S$.
For the first case, for each $e\in S$, we have $y^{(t+1)}_e=y^{(t)}_e$ with probability $1$, and thus, the inequality holds.
For the second case, assuming that $p\in S$, we have:
\[
\E\left[\prod_{e\in S} y^{(t+1)}_e \mid \set{y^{(t)}_e}_{e\in S} \right] = \E\left[ y^{(t+1)}_p \mid y^{(t)}_p \right]\cdot \prod_{e\in S\setminus\set{p}}y^{(t)}_e
=y^{(t)}_p\cdot\prod_{e\in S\setminus\set{p}}y^{(t)}_e~.
\]

For the third case, we have:
\begin{align*}
\E\left[\prod_{e\in S} y^{(t+1)}_e \mid \set{y^{(t)}_e}_{e\in S} \right] 
&= \E\left[ y^{(t+1)}_p\cdot X^{(t+1)}_q \mid y^{(t)}_p,y^{(t)}_q \right]\cdot \prod_{e\in S\setminus\set{p,q}}y^{(t)}_e \\
&\leq y^{(t)}_p\cdot y^{(t)}_q\prod_{e\in S\setminus\set{p,q}}y^{(t)}_e~.    
\end{align*}

\subsection{Proof of~\cref{lem:concentration}}

In this proof, we need to leverage several existing results.

\begin{lemma}[Negatively Correlated Chernoff Bound]
Suppose that $\set{X_i}_{i\in[n]}$ is negatively correlated.
Let $X:=\sum_{i\in[n]}X_i$ and $\mu:=\E[X]$.
For any $\delta\geq 0$, we have $\Pr[X\geq (1+\delta)\mu]\leq \exp(-\delta^2\mu/(2+\delta))$.
\label{def:chernoff}
\end{lemma}

Given any fractional solution $\vx\in[0,1]^{E}$, the  concave pessimistic estimator~\cite{HO14} is as follows:
\[
g_{t,\theta}(x):=e^{-\theta t} \cdot \E_{U\sim \cD(\vx)} \left[ e^{\theta \cdot f(U)} \right],
\]
where $t:=(1-\delta)F(\vx)$, $\theta:=\ln(1-\delta)<0$ and $\cD(\vx)$ stands for the product distribution (i.e., $\Pr_{U\sim \cD(\vx)}[e\in U]=x_e$ for all $e\in E$).

The following properties of the concave pessimistic estimator are shown by~\cite{HO14}.
\ref{pro:super} of \cref{lem:est} holds because $\exp(\theta\cdot f(\cdot))$ is a supermodular function if $f$ is submodular.
\ref{pro:bound} of \cref{lem:est} follows the standard proof for Chernoff bound.
\begin{lemma}[\cite{HO14}]
The concave pessimistic estimator has the following properties:
\begin{enumerate}[label=(\ref{lem:est}\ablue{\alph*}),leftmargin=*,align=left]
\item $\E_{U\sim \cD(\vx)}[\exp(\theta \cdot f(U))]$ is a multilinear extension of a supermodular function.
\label{pro:super}
\item $g_{t,\theta}(\vx)\leq \exp(-\delta^2\cdot F(\vx)/2)$.
\label{pro:bound}
\end{enumerate}
\label{lem:est}
\end{lemma}

Next, we show that the fractional solution $\vy$ computed by \cref{alg:knapsack-pipage} admits two other properties under the above concave pessimistic estimator.

\begin{lemma}
Let $\vy$ be the solution computed by \cref{alg:knapsack-pipage} at the end of the while-loop.
Then, we have:
\begin{enumerate}[label=(\ref{lem:est-knap}\ablue{\alph*}),leftmargin=*,align=left]
\item $g_{t,\theta}$ is concave along the line $(c_a \ve_a - c_b \ve_b)$ for any $a,b\in E$ and constants $c_a,c_b\geq 0$.
\label{pro:concave}
\item $\1\set{F(\vy)\leq t} \leq g_{t,\theta}(\vy)$, where $\1\set{F(\vy)\leq t}=1$ if $F(\vy)\leq t$; otherwise, $\1\set{F(\vy)\leq t}=0$. 
\label{pro:bridge}
\end{enumerate}
\label{lem:est-knap}
\end{lemma}
\begin{proof}
    To show \ref{pro:concave}, we need to prove that the function $g_{t,\theta}(\vy+c_a\lambda \ve_a - c_b \lambda \ve_b)$ is concave with respect to $\lambda$ for a fix point $\vy\in[0,1]^E$.
By \ref{pro:super} of \cref{lem:est}, we know that the function $g_{t,\theta}$ is a multilinear extension of a supermodular function multiplying a constant $e^{-\theta t}$.
By \ref{pro:partial-diff} of \cref{lem:extension-diff} and noting that the negation of a supermodular function is submodular, we know that $\frac{\partial^2 F}{\partial x_i \partial x_j}\geq 0$ if $F$ is a supermodular function's multilinear extension.
By the proof of \cref{lem:plane-convex}, we know that the second-order derivation of $g_{t,\theta}(\vy+c_a\lambda \ve_a - c_b \lambda \ve_b)$ in terms of $\lambda$ is a non-positive value.
Thus, \ref{pro:concave} holds.

To show \ref{pro:bridge}, there are two cases based on whether $\vy$ is integral or not.
If $\vy$ is integral, then $g_{t,\theta}(\vy)=\exp(\theta (f(U)-t))$ where $U:=\set{e\in E: y_e=1}$.
In this case, $F(\vy)$ is equal to $f(U)$.
If $F(\vy)\leq t$ (i.e., $f(U)\leq t$), then we need to prove $1\leq g_{t,\theta}(\vy)$ (i.e., $\exp(\theta(f(U)-t))$).
This is true because $\theta<0$; so $\exp(\theta(f(U)-t))\geq 1$.
If $F(\vy)> t$, then we need to prove $0\leq g_{t,\theta}(\vy)$.
This is always true since $\exp(\theta(f(U)-t))\geq 0$ always holds.
If $\vy$ is not integral and $F(\vy)>t$, then the claim holds since the function $g_{t,\theta}$ is always non-negative.
If $F(\vy)\leq t$, we need to prove $1\leq g_{t,\theta}(\vy)$.
By Jensen's inequality, we have
\[
e^{-\theta t} \cdot e^{\theta\E_{U\sim \cD(\vy)}[f(U)]} \leq e^{-\theta t} \cdot \E_{U\sim \cD(\vx)} \left[ e^{\theta \cdot f(U)} \right].
\]
Observe that the left-hand side is $\exp(\theta(F(\vy)-t))$, which must be larger than $1$ since $\theta<0$.
This finishes proving the claim.
\end{proof}

\begin{lemma}
Given any function $g:[0,1]^E\to \R_{\geq 0}$ such that $g$ is concave along the line $(c_a \ve_a - c_b \ve_b)$ for any $a,b\in E$ and constants $c_a, c_b$, then $\E[g(\vx)]\leq g(\vy)$, where $\vy$ is the solution computed by \cref{alg:knapsack-pipage} at the end of the while-loop.    
\label{lem:concave-alg}
\end{lemma}
\begin{proof}
    The proof is similar to the analysis in the proof of~\cref{lem:knap-pipage:obj}, in which the function (multilinear extension) is convex along the line $(c_a \ve_a - c_b \ve_b)$.
However, we need to reverse the direction of inequality since $g(\cdot)$ is concave along the line $(c_a \ve_a-c_b \ve_b)$.
Inducing the lemma over iterations gives the concave property.
\end{proof}

\begin{proof}[Proof of \cref{lem:concentration}]
Combining \cref{lem:est}, \cref{lem:est-knap}, and \cref{lem:concave-alg}, we have:
\[
\Pr[F(\vy)\leq t] \leq \E[g_{t,\theta}(\vy)] \leq g_{t,\theta}(\vx) \leq \exp(-\delta^2 F(\vx)/2).
\]
The first inequality is due to \ref{pro:bridge} of \cref{lem:est-knap}, where we take the expectation over the distribution produced by \cref{alg:knapsack-pipage} on both sides.
The second inequality is due to \cref{lem:concave-alg}.
The third inequality is due to \ref{pro:bound} of \cref{lem:est}.
\end{proof}

\end{document}